\documentclass[fleqn,usenatbib]{mnras}

\usepackage{newtxtext,newtxmath}
\usepackage{xcolor}

\usepackage[T1]{fontenc}

\DeclareRobustCommand{\VAN}[3]{#2}
\let\VANthebibliography\thebibliography
\def\thebibliography{\DeclareRobustCommand{\VAN}[3]{##3}\VANthebibliography}

\usepackage{graphicx}	
\usepackage{amsmath}	
\usepackage{lipsum}
\usepackage{float}

\title[$^{26}$Al and $^{60}$Fe in the LMC]{Evolution of radioactive elements in the LMC: predictions for future $\gamma$-ray surveys}

\author[A. Vasini et al.]{
Arianna Vasini,$^{1,2}$\thanks{E-mail: arianna.vasini@inaf.it}
F. Matteucci,$^{1,2,3}$, E. Spitoni$^{2,4}$ and Thomas Siegert$^{5}$
\\
$^{1}$Dipartimento di Fisica, Sezione di Astronomia, Università di Trieste, via G.B. Tiepolo 11, I-34131, Trieste, Italy\\
$^{2}$I.N.A.F. Osservatorio Astronomico di Trieste, via G.B. Tiepolo 11, I-34131, Trieste, Italy\\
$^{3}$I.N.F.N. Sezione di Trieste, Via Valerio 2, 34127, Trieste, Italy\\
$^{4}$ Universit\'e C$\hat{o}$te d’Azur, Observatoire de la C$\hat{o}$te d’Azur, CNRS, Laboratoire Lagrange, Bd de l’Observatoire, CS 34229,
06304 Nice Cedex 4, France\\
$^{5}$ Julius-Maximilians-Universität Würzburg, Fakultät für Physik und Astronomie, Institut für Theoretische Physik und Astrophysik, \\Lehrstuhl für Astronomie, Emil-Fischer-Str. 31, D-97074 Würzburg, Germany
}

\date{Accepted XXX. Received YYY; in original form ZZZ}

\pubyear{2015}

\begin{document}
\label{firstpage}
\pagerange{\pageref{firstpage}--\pageref{lastpage}}
\maketitle

\begin{abstract}
Short-lived radionuclides, such as $^{26}$Al and $^{60}$Fe, are tracers of star formation. Therefore, their abundances can unravel the recent star formation history of the host galaxy. In view of future $\gamma$-ray surveys, we predict the masses and fluxes of these two elements in the Large Magellanic Cloud (LMC) using new chemical evolution models. Our best model reproduces the abundance patterns of [$\alpha$/Fe] ratios, the gas mass, the average metallicity, the present time supernova and nova rates observed in LMC. We show three main results: i) the best model for the LMC suggests a star formation rate very mild at the beginning with a recent burst, and a Salpeter-like initial mass function. ii) The predicted mass of $^{26}$Al is 0.33 M$_{\odot}$, 2/3 produced by massive stars and 1/3 by novae. iii) The predicted mass of $^{60}$Fe is 0.44 M$_{\odot}$, entirely produced by massive stars. This result suggests a larger fraction of $^{60}$Fe, at variance with the Milky Way. The explanation for this lies in the adopted initial mass function, that for the LMC contains more massive stars than for the Milky Way. These predictions can be useful for the COSI-SMEX mission planned for launch in 2027. The expected $\gamma$-ray line fluxes for the 1.809 MeV line of $^{26}$Al and the 1.173 and 1.332 MeV lines of $^{60}$Fe are in the range of $(0.2$--$2.7)\times 10^{-6}\,\mathrm{ph\,cm^{-2}\,s^{-1}}$ and $(0.7$--$2.8)\times 10^{-7}\,\mathrm{ph\,cm^{-2}\,s^{-1}}$, respectively. This new instrument could have the sensitivity to detect the upper end of the predicted 1.8 MeV flux within its nominal two-year mission.
\end{abstract}




\section{Introduction}

Only a few chemical evolution papers on the evolution of the abundances of radioactive isotopes have appeared in the last years, due to the additional difficulty connected to their decay, even though they could be a precious opportunity to add knowledge to the present time chemical scenario. In this study we focus on two specific isotopes, $^{26}$Al and $^{60}$Fe, that share similar characteristics in terms of mean lifetime and stellar progenitors. They both decay on a timescale of the order of $\sim$ Myr, in particular they have decay time $\tau_{26Al}$=1.05 Myr and $\tau_{60Fe}$=3.75 Myr respectively (for references see \citealt{Diehl13}). This timescale, from an astrophysical point of view, is short, hence the label of short-lived radioactive isotopes. Regarding the stellar producers, these two isotopes are both synthesized by massive stars (M $\ge$ 13 M$_{\odot}$) with an additional contribution, only for $^{26}$Al, coming from nova systems. The direct consequence of these two features combined together is that the abundances of $^{26}$Al and $^{60}$Fe do not contain any information regarding the early times of their host galaxy but depend only on its recent history, thus being precise tracers of active star formation.

Regarding the observational aspect, measurements are available only for the Milky Way  \citep{Diehl+95,Diehl13,Kretschmer+13,Pleintinger+20} whereas for other nearby structures these isotopes have never been observed. The lack of data is partly due to the techniques needed for the detection, that are completely different from those applied to observe the stable element abundances. For unstable isotopes the detection is based on $\gamma$-astronomy techniques, in particular on the $\gamma$-ray spectroscopy. When $^{26}$Al and $^{60}$Fe decay they produce $\gamma$ photons, one for $^{26}$Al at 1.809 MeV and two for $^{60}$Fe at 1.173 MeV and 1.332 MeV. The normal spectrographs are not sensitive to this energy band, hence dedicated instruments must be built, targeting specifically these two nuclei. Things could change in the future since NASA is working on a new $\gamma$ survey that could also target the Large Magellanic Cloud (LMC), crossing the boundaries of the Galaxy for the first time. 
The currently best $\gamma$-ray spectrometer telescope, SPI \citep{Vedrenne+03} onboard ESA's INTEGRAL satellite \citep{Winkler+03} revolutionised our understanding of radioactive isotopes, star formation \citep{Diehl+06}, and superbubble structure of the interstellar medium \citep{Krause+15} through its unprecedented spectral resolution. Due to the large instrumental background in the MeV photon range, strongly limiting the sensitivity, SPI cannot\footnote{It would require on the order of 6 years of dead-time corrected observation time to observe the LMC with SPI.} observe the LMC in the light of decaying nuclei. NASA's new Compton Spectrometer and Imager (COSI, \citealt{Tomsick+19}), slated for launch in 2027, will have an improved sensitivity over SPI and will be working as a survey instrument for $\gamma$-ray lines and continuum sources. The LMC could then be the first extragalactic source for which radioactive isotopes could be measured directly through their $\gamma$-ray signatures.

In parallel with the observations it is thus necessary to develop suitable chemical evolution models to provide theoretical constraints to the upcoming observations. Unfortunately, in the literature just few examples can be found. The first attempts were performed by \citet{Clayton84} and \citet{Clayton88} with analytical models. The problems related to analytical models lie in the several approximations necessary to perform the calculations, the first one being the instantaneous recycling approximation (IRA) that assumes that stars smaller than the Sun never die and those more massive die right after their birth, re-injecting their nucleosynthesis products immediately. Later on, \citet{Timmes+95} carried out the first study on radioactive isotopes calculating explicitly the amount of $^{26}$Al and $^{60}$Fe throughout the Milky Way with a numerical model. Recently, \citet{Vasini+22} refined these calculations with a more elaborated and detailed numerical model and predicted the total masses of $^{26}$Al and $^{60}$Fe, concluding that novae must have an important role in the $^{26}$Al production.

This paper is aimed at computing a model for the chemical evolution of the LMC that reproduces the main observational features of this galaxy, including  the [$\alpha$/Fe] versus [Fe/H] relations, as well as to provide the  first theoretical estimates of $^{26}$Al and $^{60}$Fe masses. The model takes into account detailed stellar lifetimes, yields and SN progenitors and is able to follow the chemical evolution of 30 either stable or radioactive isotopes at the same time. The type of stars considered in the calculations are AGB stars, core-collapse supernovae (cc-SNe), Type Ia supernovae (SNe Ia) and nova systems. Our main purpose is to offer predictions about the mass of $^{26}$Al and $^{60}$Fe in the framework of a good model for LMC, thus offering theoretical constraints in terms of $\gamma$-ray line fluxes to the future $\gamma$-ray survey mission COSI \citep{Tomsick+19}.
The paper is organized as follows: in Section~\ref{sec:model} we present the chemical evolution model together with all the adopted parameters, in Section~\ref{sec:results} we show the results relative to $^{26}$Al and $^{60}$Fe masses as well as the $\alpha$-element patterns. In Section~\ref{sec:dicussion} we discuss the results and the constraints that we derive on the evolution of the LMC and predicted the $\gamma$-ray line fluxes, and finally in Section~\ref{sec:conclusions} we summarize the work and we draw the main conclusions.

\section{The Model}
\label{sec:model}
In this Section we show the model used to compute the evolution of the LMC and in particular of both stable and unstable elements ($^{26}$Al and $^{60}$Fe). We describe the fundamental equations governing the evolution of the chemical abundances in the interstellar medium (ISM) (Section ~\ref{sec:CEEq}). Since we performed four tests, with different star formation histories, we show all the common assumptions in Section~\ref{sec:maths} and separately each star formation rate (SFR) with its initial mass function (IMF) in Section~\ref{sec:SFR}. Last, we discuss the yields both for stable and for unstable nuclei in Section~\ref{sec:yields}.

\subsection{Chemical Evolution Equation}
\label{sec:CEEq}

The chemical evolution model we used to simulate the LMC is based on a chemical evolution equation whose versions are largely discussed in the literature. In particular, we will refer to that presented by \citet{Vasini+22} with an additional term related to the wind contribution:
\begin{equation}\begin{split}
\frac{d(X_{\rm i}M_{\rm gas})}{dt}=&-X_{\rm i}(t)\psi (t) -\lambda_{\rm i}X_{\rm i}(t)M_{\rm gas}- X_{\rm i}(t)W_{\rm i}(t)\\
& +R_{\rm i}(t)+X_{\rm i,A}A(t)
    \label{eq:CEeq}
\end{split}\end{equation}

This equation describes how the amount of the element $i$ in the ISM changes in time under the effect of several phenomena, which act either increasing or depleting the abundance of the element $i$ itself. The depletion of the ISM comes from the firsts three terms, $-X_{\rm i}(t)\psi (t)$, $-\lambda_{\rm i}X_{\rm i}(t)M_{\rm gas}$ and $- X{\rm i}(t)W_{\rm i}(t)$, where M$_{\rm gas}$ is the mass of gas and X$_{\rm i}$ is the abundance by mass of the element $i$. In order, these terms represent the rate of inclusion of gas into new stars, the radioactive decay of the unstable isotopes in the ISM and the outflow of the same gas due to the galactic wind. On the other side, $R_{\rm i}(t)$ and $X_{\rm i,A}A(t)$ account for the evolution of abundances by the injection of nucleosynthetic products into the ISM at the moment of the stellar explosion and by the infall from extragalactic gas.

Regarding the radioactive decay and its corresponding term, $\lambda_{\rm i}$ is the inverse of the decay time scale $\tau_{\rm i}$ of the element $i$, which, for the case of $^{26}$Al and $^{60}$Fe, is equal to $\tau_{\rm 26Al}$=1.05 Myr and $\tau_{\rm 60Fe}$=3.75 Myr. It is also important to point out that Equation~\ref{eq:CEeq} is a general version useful to treat both stable and unstable nuclei. Indeed, nuclear physics states that the decay time $\tau$ of a stable isotope tends to infinity and therefore its inverse $\lambda$ tends to zero. Therefore, if we assume $\lambda_{\rm i}$=0, the radioactive decay term cancels and Equation~\ref{eq:CEeq} becomes the classic chemical evolution equation for stable isotopes.
\newline

\subsection{Model assumptions}
\label{sec:maths}
The chemical evolution model we used is a one-zone model, which means that we assume the LMC to be a structure with a homogeneous chemical composition, and adopt a one-infall process for galaxy formation that follows the equation: 
\begin{equation}
    A(t) = a e^{-t/\tau_{\rm LMC}}
    \label{eq:one_infall}
\end{equation}
where we set $\tau_{\rm LMC}$=5 Gyr as the time-scale of the infall, and $a$ is one of the adopted parameter obtained by reproducing the present time total mass of the LMC. It is derived by imposing an infall mass that should be reached at the end of the accretion process. Here we assume for all models $M_{\rm inf}=6\cdot 10^{9}M_{\odot}$.

One of the most important parameter in the Equation~\ref{eq:CEeq} is the SFR, $\psi(t)$. The general formulation of the star formation rate we adopted is the one by \citet{Kennicutt1998}:
\begin{equation}
    \psi(t) \propto \nu M_{\rm gas}^{k}(t)
    \label{eq:SK}
\end{equation}
where the SFR is proportional to the mass of the gas through the factor $\nu$, which represents the efficiency of star formation. In this study we tested different cases with different values of $\nu$, both constant in time and time-dependent that will be presented in detail in Section~\ref{sec:SFR}. 

To reproduce faithfully the LMC we also accounted for the contribution by galactic winds $W_{\rm i}(t)$, which is proportional to the SFR according to:

\begin{equation}
W_{\rm i}(t)=\omega_{\rm i}\psi(t)
\label{eq:wind}
\end{equation}
where $\omega_{\rm i}$, the mass loading factor, represents the wind efficiency and depends on the element analysed, and $\psi(t)$ is the SFR. In our case, we assume that the mass loading factor is $\omega$=0.25 Gyr$^{-1}$ (see \citealt{Bradamante+98}) for every element.
The wind acts on the chemistry of the galaxy by subtracting gas and therefore lowering the amount of stars that can be formed and, consequently, the amount of metals that are injected into the ISM. The net effect of the wind is, basically, lowering the efficiency of star formation $\nu$. The wind enters in Equation~\ref{eq:CEeq} with the term $X_{\rm i}(t)W_{\rm i}(t)$.

Moving on to the ISM polluters, our model accounts for several types of stars which explode according to different scenarios, each of which is implemented in the model by computing its specific rate. Starting from SNe Ia we assume the rate suggested by \citet{Matteucci&Greggio86} and \citet{Matteucci&Recchi01}, where details can be found, relative to the single degenerate scenario:
\begin{equation}
R_{\rm SNIa}(t)=A_{\rm B}\int^{M_{\rm BM}}_{M_{\rm Bm}} \varphi(m)\bigg[\int^{0.5}_{\mu_{\rm Bmin}} f(\mu_{\rm B})\phi(t-\tau_{\rm m2})\bigg]dm
    \label{eq:SNIa}
\end{equation}
where A$_{\rm B}$=0.035 is the fraction of binary systems that can give rise to a Type Ia SNa and is derived by reproducing the present time observed rate. The mass $m$ over which the integration is performed, represents physically the sum of the masses of the two companions and can span from $M_{\rm Bm}$ (smallest total mass of the system) to $M_{\rm BM}$ (largest total mass of the system). $M_{\rm Bm}$ is assumed to be 3 M$_{\odot}$ so that the two companions both have the minimum mass to make the white dwarf reach the Chandrasekhar mass, whereas $M_{\rm BM}$ is 16 M$_{\odot}$ since the upper limit to generate a CO white dwarf is 8 M$_{\odot}$. $f(\mu_{\rm B})$ is the distribution of the mass ratio of the two system components and $\tau_{\rm m2}$ is the life time of the smallest one, which dies later and therefore represents the clock of the whole system (see \citealt{Matteucci&Recchi01} for details).

The cc-SN rate is:
\begin{equation}
R_{\rm SNII,Ib,Ic}(t)=\int^{M_{\rm U}}_{M_{\rm L}} \psi(t-\tau_{\rm m})\varphi(m)dm
    \label{eq:CC-SN}
\end{equation}
with $\tau_{\rm m}$ being the stellar lifetime, $M_{\rm U}$=30 M$_{\odot}$, $M_{\rm L}$=8 M$_{\odot}$ in the case of Type II SN, $M_{\rm U}$=100 M$_{\odot}$, $M_{\rm L}$=30 M$_{\odot}$ for Type Ib and Type Ic SN. 

Finally, considering that an important contribution to $^{26}$Al is given by nova systems, as suggested by \citet{LC06} and \citet{Vasini+22}, we also assumed the contribution by the nova system outbursts as done by \citet{D'antona&Matteucci91}:
\begin{equation}
R_{\rm novaout}(t)=10^{4}\cdot\alpha\int^{8}_{0.8}\psi(t-\tau_{\rm m}-\Delta t)\varphi(m)dm
\label{eq:nove}
\end{equation}
A nova system can produce several outbursts during its whole lifetime. Therefore, to consider properly this contribution we have to account for the rate of formation of nova systems, and the number of outbursts of each system, expressed in Equation~\ref{eq:nove} by the factor $10^{4}$ \citep{Ford+78}. $\alpha$ represents the fraction of binary systems that can actually give rise to a nova system, and can be fine tuned to reproduce the observed present time nova rate. For the LMC, the nova rate measured by \citet{Mroz+16} is 2.4$\pm$0.8 events yr$^{-1}$. $\tau_{\rm m}$ is once again the stellar lifetime and $\Delta t$ is the time necessary to form a nova system and is $\ge$ 1 Gyr.

\subsection{Star Formation History}
\label{sec:SFR} 
The star formation history (SFH) is given by the combination of a SFR with an IMF. The gas depletion induced by the galactic wind, as explained in Section~\ref{sec:CEEq}, affects the SFH. As already anticipated, we tested four cases that differ for the SFH adopted. The assumed SFRs follow always Equation~\ref{eq:SK} but with a variable efficiency $\nu$ in order to reproduce the SFRs proposed in the literature.  Regarding the IMF, we adopted the one suggested in the various SFHs.
The SFHs tested are taken from \citet{Kennicutt1998}, \citet{Hasselquist2021}, \citet{HZ09} and \citet{Calura+03} and are described in the next subsections.
The choice of testing more than one SFH is due to the fact that in the literature many SFHs have been proposed but none of them  can firmly exclude the others.
Table~\ref{tab:models} summarizes the parameters of each SFH case. The first column lists the parameters of the model and the columns from the second to the fifth report the value of the parameters for the four different cases explored. In particular, the adopted parameters are: the SFR, IMF, the fraction of binary systems becoming Type Ia SNe ($A_B$), the wind efficiency ($\omega$), the infall timescale ($\tau_{\rm LMC}$), the infall mass ($M_{\rm inf}$) and the efficiency of star formation ($\nu$).
\begin{table*}
\centering
\begin{tabular}{|c|c|c|c|c|}

     \hline
      & & & & \\
     \textbf{Parameter} & \textbf{Model 1} & \textbf{Model 2} & \textbf{Model 3} &\textbf{Model 4} \\
     & & & & \\
     \hline
     SFR & \citet{Kennicutt1998} & \citet{Hasselquist2021} & \citet{HZ09} & \citet{Calura+03} \\
     \hline
     IMF & \citet{Kroupa1993} & \citet{Kroupa01} & \citet{Salpeter55} & \citet{Salpeter55}\\
     \hline
     fraction of SNIa & 0.035 & 0.035 & 0.09 & 0.09\\
     \hline
     $\omega$ [Gyr$^{-1}$] & 0.25 & 0.25 & 0.25 & 0.25\\
     (wind efficiency) & & & &\\
     \hline
     $\tau_{LMC}$ [Gyr] & 5.0 & 5.0 & 5.0 & 5.0\\
     (infall time) & & & &\\
     \hline
     M$_{inf}$ [M$_{\odot}$] & 6$\cdot$10$^{9}$ & 6$\cdot$10$^{9}$ & 6$\cdot$10$^{9}$ & 6$\cdot$10$^{9}$ \\
     \hline
     $\nu$  [Gyr$^{-1}$] & 0.4 & 0.05 (t $<$ 11 ) &  0.03 (t$<$10.8 ) & 0.05 (t$<$2 )\\
     (SF efficiency) & & 0.7 (t$>$11 ) & 0.25 (10.8 $<$ t $<$ 11.8) & 0.6 (2 $<$ t $<$ 4)\\
     & & & 0.1 (11.8 $<$ t $<$ 13.2 ) & 0.05 (4 $<$ t $<$ 12)\\
     & & & 0.4 (13.2 $<$ t $<$ 13.4 ) & 0.5 (t $>$ 12)\\
     & & & 0.35 (13.4 $<$ t $<$ 13.5 ) &\\
     & & & 0.8 (t $>$ 13.5 ) &\\
     \hline
\end{tabular}
\caption{List of parameters adopted for the four models tested. The first column contains the parameters and each of the following four columns is representative of a specific model. Notice that the differences among the models lay in the SFR adopted (first row) which is obtained fine tuning the star formation efficiency $\nu$, which is variable as well (last row).}
\label{tab:models}
\end{table*}
\begin{figure*}
    \centering
    \includegraphics[scale=0.6]{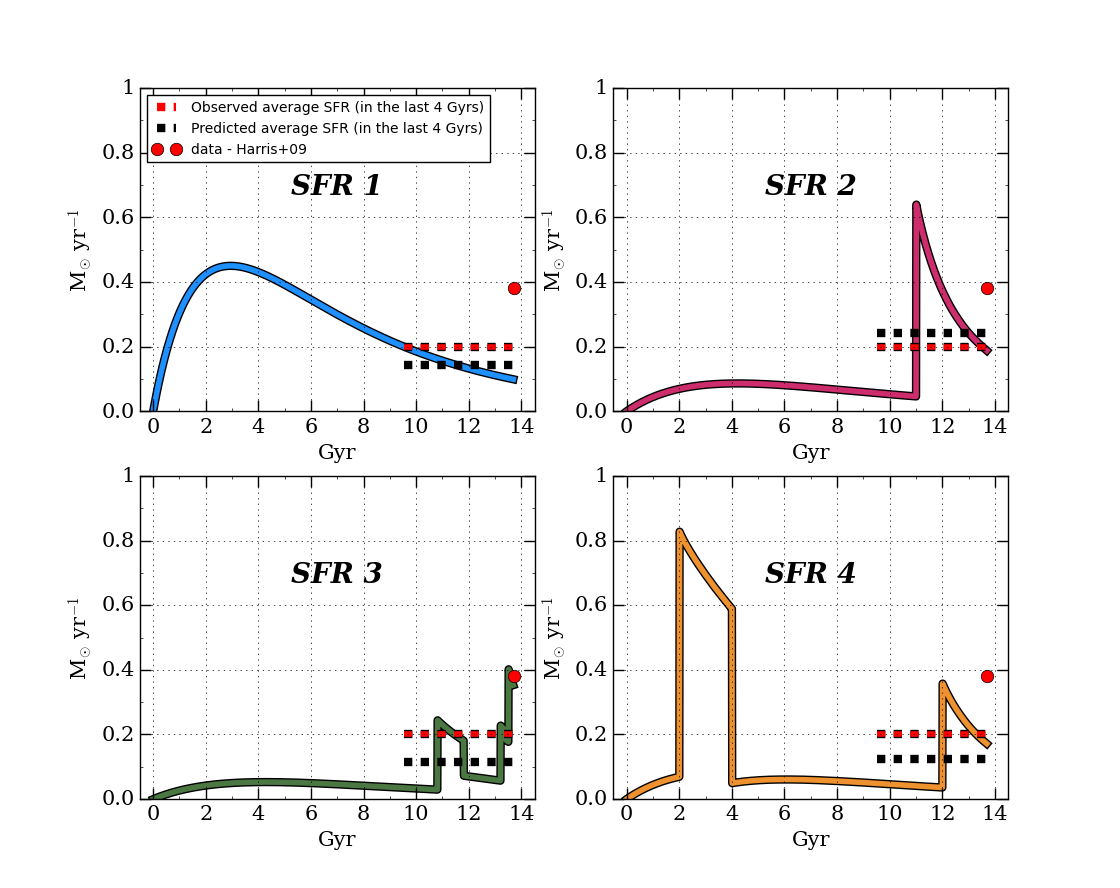}
    \caption{The four SFHs tested in this study. \textit{Top right panel}: Model 1 with a Schmidt-Kennicutt law taken from \citet{Kennicutt1998} with constant star formation efficiency $\nu$. The red point is the present time observed star formation rate, 0.38 M$_{\odot}$ yr$^{-1}$, taken from \citet{HZ09}. The dashed red line is the average SFR in the last 4.5 Gyrs, again by \citet{HZ09} and the dashed black line is the average SFR in the last 4.5 Gyrs that we obtained. \textit{Top left panel}: Model 2 with Hasselquist-like SFH inspired by \citet{Hasselquist2021}.  \textit{Bottom left}: SFH by \citet{HZ09} used for Model 3. The authors propose a SFH very low for the first 10 Gyr followed by four bursts, whereas we adopt just three bursts, since the two most recent ones are very close in time and the star formation in between is not too low. With this SFH we reproduce the observed value. \textit{Bottom right panel}: SFH by \citet{Calura+03} tested in Model 4. The characteristics of this SFH were chosen because they agreed with \citet{Bradamante+98} on purely theoretical basis but it was suitable to fit the data available at the time. Notice that only Model 3 can reproduce  the present time value and also the average SFR is consistent with the observation, given the uncertainties.} 
    \label{fig:4sfr}
\end{figure*}

\subsubsection{Model 1 - Kennicutt et al. (1998) SFR with Kroupa et al (1993) IMF}
\label{sec:SK98}
As a first case we tested the classical Schmidt-Kennicutt law, proposed by \citet{Kennicutt1998}. The mathematical form of this SFR is shown in Equation~\ref{eq:SK}. The adopted efficiency of star formation is $\nu$=0.4 Gyr$^{-1}$. 
In the top right panel of Fig.~\ref{fig:4sfr} we compare this theoretical SFR with the present time observed value and the observed average SFR in the last 4.5 Gyrs, both from \citet{HZ09}. They proposed a present time SFR in the LMC of 0.38 M$_{\odot}$ yr$^{-1}$ and an average value in the last 4.5 Gyrs of 0.2 M$_{\odot}$ yr$^{-1}$. The model predicts a lower value, but given the large uncertainties 
that affect the observations, we can consider the model acceptable since it agrees within a factor of 2 with the observation. It is worth noting that the following models are developed starting from the same Schmidt-Kennicutt law, but with a time-dependent star formation efficiency in order to reproduce the bursts of star formation. In this way we can preserve the dependence from the gas mass only by changing the efficiency $\nu$ in Equation~\ref{eq:SK}.

The IMF adopted in this case is from \citet{Kroupa1993} that we assumed as a default choice since it was derived to reproduce the features of the solar neighborhood. Mathematically:
\begin{equation}
\varphi_{\rm K93}(M)\propto \left \{ \begin{array}{rl}
M^{-1.3}\,\,\,\,\,\,\,\,\,\,\,\,\,\,\,\,\,\,\,\,\,\,\,\,\,\,\,\,\,\,\,M<0.5 M_{\odot}\\
M^{-2.2}\,\,\,\,\,\,\,\,\,\,\,\,\,\,\,\,\,\,0.5<M/M_{\odot}<1\\
M^{-2.7}\,\,\,\,\,\,\,\,\,\,\,\,\,\,\,\,\,\,\,\,\,\,\,\,\,\,\,\,\,\,\,\,\,\,\,\,M>1 M_{\odot}
\end{array}
\right.
\label{eq:kroupa93}
\end{equation}
which is valid in the interval 0.1-100 M$_{\odot}$. 

\subsubsection{Model 2 - step function and Kroupa 2001 IMF}
The second SFR we tested is a step function that was modeled following the theoretical one by \citet{Hasselquist2021}. This SFH suggests a low and almost constant value for the SFR until 2 Gyrs ago followed by a burst. The present time predicted SFR fits nicely the observed value proposed by \citet{HZ09}.

The IMF we assumed in this case is the one that also \citet{Hasselquist2021} used, namely the two-slope IMF from \citet{Kroupa01}, very similar in shape to a \citet{Salpeter55} law for masses $M> 0.5M_{\odot}$. From this IMF we expect a larger metal production because the contribution by massive stars is higher than in the \citet{Kroupa1993} previously introduced. The mathematical form of this IMF is the following:
\begin{equation}
\varphi_{\rm K01}(M)\propto \left \{ \begin{array}{rl}
M^{-1.3}\,\,\,\,\,\,\,\,\,\,\,\,\,\,\,\,\,\,M<0.5 M_{\odot}\\
M^{-2.3}\,\,\,\,\,\,\,\,\,\,\,\,\,\,\,\,\,\,M>0.5 M_{\odot}
\end{array}
\right.
\label{eq:kroupa01}
\end{equation}
again valid between 0.1-100 M$_{\odot}$.

\subsubsection{Model 3 - Harris \& Zaritsky (2009) SFR with Salpeter (1995) IMF}
\label{sec:HZ09}
The third presented model is that by \citet{HZ09} where a more complex SFR is suggested. The shape they propose, presented in their Fig. 11, is obtained by collecting the color-magnitude diagram (CMD) of the LMC divided in a sub-grid. For each portion of the grid many synthetic CMDs were produced assuming different SFHs and then the best one was selected. Lastly, they performed a convolution between the SFHs around the grid and obtained the final one.
Their SFH is quite similar to the \citet{Hasselquist2021} one since also in this case it is flat for the first $\sim$ 11 Gyr. Then, there are four peaks very close in time which are consistently higher than the early SFR.
In the bottom right panel of Fig.~\ref{fig:4sfr} we show how we modeled this third SFH, again by means of the  Schmidt-Kennicutt law with a non-constant star formation efficiency. We highlight that we merged the two most recent bursts in one single burst due to the fact that they happened very close to each other.

The IMF we adopted for Model 3 is the one by \citet{Salpeter55} as adopted by \citet{HZ09}:
\begin{equation}
\varphi_{\rm S}(M)\propto M^{-2.35}
\end{equation}
valid in the mass range 0.1-100 M$_{\odot}$.

\subsubsection{Model 4 - Calura et al. (2003) and Salpeter (1955) IMF}
\label{sec:calura}
The last model we tested is the SFH proposed by \citet{Calura+03}. In that paper it was proposed a double burst scenario, with the first one between 2 and 4 Gyr and the second one, less intense, which started at around 12 Gyr and is still going on with a present time value that reproduces quite well the observations, as shown in the bottom left panel of Fig.~\ref{fig:4sfr}.

The assumed IMF was the \citet{Salpeter55} one.

\subsection{Nucleosynthesis and yields}
\label{sec:yields}

Concerning the nucleosynthesis, our model accounts for the contribution by different types of stars as massive stars, AGB stars, SNIa and nova systems. For each type of star we assumed the state-of-the-art yields.  

In the case of stable isotopes, for massive stars we assumed yields by \citet{Kobayashi+06}, for AGBs we adopted those by \citet{Karakas10} and for SNIa we used the yields by \citet{Iwamoto+99}. Regarding the contribution by nova systems we adopted the nucleosynthesis proposed by \citet{JH98} making an assumption on the nova chemical composition. As done in \citet{Romano&Matteucci03}, we assume that the 30\% of nova systems are ONe novae and the remaining 70\% are CO novae, hence the final yield we used is a weighted average of those proposed by \citet{JH98}. 

For what concerns the two radioactive isotopes, $^{26}$Al and $^{60}$Fe, for AGB stars and Type Ia SNe we assumed the same sets of yields adopted for the stable elements. For massive stars we adopted the yields by \citet{WW95}. Finally, for the production of $^{26}$Al by nova systems we performed a first test adopting the yield by \citet{JH07}, and a following test excluding the nova production, to double check how much they do contribute. The choice of using the set by \citet{JH07} is justified by their adoption in \citet{Vasini+22} where it is shown that this set is the best one to reproduce the Milky Way observations.

More recent yields for massive stars appeared in the literature, such as \citet{Kobayashi+11} and \citet{LC18} but we have decided to adopt the yields by \citet{WW95} for comparison with our previous work \citep{Vasini+22}. In addition, the yields of $\alpha$-elements by \citet{Kobayashi+11} have not changed from the previous paper \citep{Kobayashi+06} that we are adopting in this work. Concerning the yields by \citet{LC06} and \citet{LC18}, we have already tested them for the Milky Way and they were providing a worse fit to the $^{26}$Al and $^{60}$Fe masses than \citet{WW95} ones.

Notice that the nuclear decay inside the stars was computed by means of an exponential factor acting on the yields of $^{26}$Al and $^{60}$Fe ejected by the stars, and depending on the stellar lifetime.

\section{Results}
\label{sec:results}
In this Section we show the results we obtained with the four SFH cases explored. We discuss the observational constraint used to select the best model and then the results obtained with that one. 

\subsection{Observational constraints}
\label{sec:obs}
In order to chose our best model, we have to ensure that as many observational constraints as possible are reproduced within their error bars, also considering that the chemical 
evolution model itself has several sources of uncertainties. Among these, the most important is related to the uncertainties in the nucleosynthesis yields.
The observational constraints are shown in Table~\ref{tab:constraints}. The first column indicates the observable quantity considered, the second indicates the label of the model, the third shows the predicted value, the fourth lists the observed value taken from the literature and the fifth reports the reference in the literature from which the observed value is taken. The observed and theoretical quantities reported in  Table~\ref{tab:constraints} are: present time SFR, present time SN rates (CC+Ia), nova rate, gas mass, stellar mass and average metallicity in LMC. It is worth noting that in the third column, about the present time nova rate, two values are reported. In Equation~\ref{eq:nove} indeed the factor $\alpha$ is a free parameter of the model and its value can change from galaxy to galaxy. In this case we considered two possible values of $\alpha$, 0.0115 and 0.024, and the two values in the table are obtained using these two prescriptions. The best $\alpha$ is chosen in order to reproduce the present time nova outburst rate.

According to this table, the two best models are Model 3 and Model 4. It is worth noting that the final total mass (gas plus stars dead and alive) agrees with the assumed infall mass for all the models.
Looking at all the other quantities, we can analyse the predictions by Model 1; for SNe rate, nova rate, gas mass and metallicity Z, the values are not in good agreement with 
the observations, hence we can easily exclude Model 1. 
Regarding Model 2, we can note that the predictions about the nova rate and the metallicity Z are higher than the expectations. These two quantities are rather important for the study we are carrying on, so we can rule out also Model 2 since it does not fit these two observables. 

\begin{table*}
\centering
\begin{tabular}{crccc}
       \hline
       \textbf{Observable} & \textbf{Model \#} & \textbf{Predicted value} & \textbf{Observed value} & \textbf{References} \\
       \hline
       Present time SFR & 1 & 0.1 M$_{\odot}$\,\,yr$^{-1}$ & $0.2_{-0.05}^{+0.6}$ M$_{\odot}$\,\,yr$^{-1}$ & \citet{HZ09}\\
       & 2 & 0.2 M$_{\odot}$\,\,yr$^{-1}$ & & \\
       & 3 & 0.35 M$_{\odot}$\,\,yr$^{-1}$ & & \\
       & 4 & 0.17 M$_{\odot}$\,\,yr$^{-1}$ & & \\
       \hline
       SNCC+SNIa rate & 1 & 0.064 SN\,\,century$^{-1}$ & 0.67\,$^{+0.49}_{-0.35}$\, SN\,\,century$^{-1}$ & \citet{Mannucci+05} \\
       & 2 & 0.24 SN\,\,century$^{-1}$  & & \\
       & 3 & 0.28 SN\,\,century$^{-1}$  & 0.86$\pm$0.25 SN\,\,century$^{-1}$ & \citet{Maoz+10} \\
       & 4 & 0.15 SN\,\,century$^{-1}$  & & \\
       \hline
       Nova rate & 1 &  5.66-11.82\,novae\,\,yr$^{-1}$ & 2.4$\pm$0.8\, novae\,\,$yr^{-1}$ & \citet{Mroz+16} \\
       & 2 &  3.9-8.2\,novae\,\,yr$^{-1}$ & & \\
       & 3 &  1.27-2.7\,novae\,\,yr$^{-1}$ & & \\
       & 4 &  2.1-4.3\,novae\,\,yr$^{-1}$ & & \\
       \hline
       Present time gas mass & 1 & 2.5$\cdot$10$^{8}$ M$_{\odot}$ & 5$\cdot$10$^{8}$ M$_{\odot}$ (neutral\,gas) & \color{blue}van der Marel \color{black}(\color{blue}2006\color{black}) \\
       & 2 &  2.7$\cdot$10$^{8}$ M$_{\odot}$ & 7$\cdot$10$^{8}$ M$_{\odot}$ & \citet{MC83}\\
       & 3 &  4.4$\cdot$10$^{8}$ M$_{\odot}$ & & \\
       & 4 &  3.4$\cdot$10$^{8}$ M$_{\odot}$ & & \\
       \hline
       Present time stellar mass & 1 &  2.28$\cdot$10$^{9}$ M$_{\odot}$ & 2.7$\cdot$10$^{9}$ M$_{\odot}$ & \color{blue}van der Marel \color{black}(\color{blue}2006\color{black}) \\
       & 2 &  0.9$\cdot$10$^{9}$ M$_{\odot}$  &  & \\
       & 3 &  0.6$\cdot$10$^{9}$ M$_{\odot}$  &  & \\
       & 4 &  1.5$\cdot$10$^{9}$ M$_{\odot}$  &  & \\
       \hline
       Present time total mass & 1 &  5.2$\cdot$10$^{9}$\,M$_{\odot}$ & 3.6$\cdot$10$^{9}$\,M$_{\odot}$ & \color{blue}van der Marel \color{black}(\color{blue}2006\color{black}) \\
       & 2 &  5.2$\cdot$10$^{9}$\,M$_{\odot}$ & & \\
       & 3 &  5.2$\cdot$10$^{9}$\,M$_{\odot}$ & & \\
       & 4 &  5.2$\cdot$10$^{9}$\,M$_{\odot}$ & & \\
       \hline
       Present time metallicity & 1 &  Z = 0.019  & 0.3-0.5\,\,Z$_{\odot}$ = 0.006 - 0.01 & \citet{Nosowitz+22} \\
       & 2 &  Z = 0.05  & Z = 0.01 (gas metallicity) & \citet{Tchernyshyov+15} \\
       & 3 &  Z = 0.015 & & \\
       & 4 &  Z = 0.024 & & \\
       & & & & \\
       & 1 &  [Fe/H] = 0.18  & [Fe/H]=-0.3 dex (upper limit) & \citet{Hasselquist2021} \\
       & 2 &  [Fe/H] = 0.38  & [Fe/H]=-0.2 dex (upper limit) & \citet{Nidever+20}\\
       & 3 &  [Fe/H] = 0.016  & & \\
       & 4 &  [Fe/H] = 0.29  & & \\
       \hline
\end{tabular}
\caption{Observational quantities used to constrain the chemical evolution model. The first column indicates the quantity considered, the second one lists the model that predicted the value in the following column, the third one contains the four values predicted by the four different models, the fourth shows the observed value with the related uncertainty when provided and the fifth reports the reference in the literature where the observed value was taken. It is highlighted that Model 1 and Model 2 do not properly reproduce the present time observational constraint, therefore they are excluded from further considerations. Regarding Model 3 and Model 4, relying only on this table, is not possible to prefer one over the other since both fail in reproducing some constraints.}
\label{tab:constraints}
\end{table*}

Regarding Model 3 and Model 4 (\citealt{HZ09} and \citealt{Calura+03} respectively) a deeper discussion is needed. Looking at Table~\ref{tab:constraints}, the present time SFR and stellar mass are more in agreement in Model 4, whereas the SNe rate, present time gas mass and metallicity Z favours Model 3. 

\begin{figure*}
    \centering
    \includegraphics[scale=0.6]{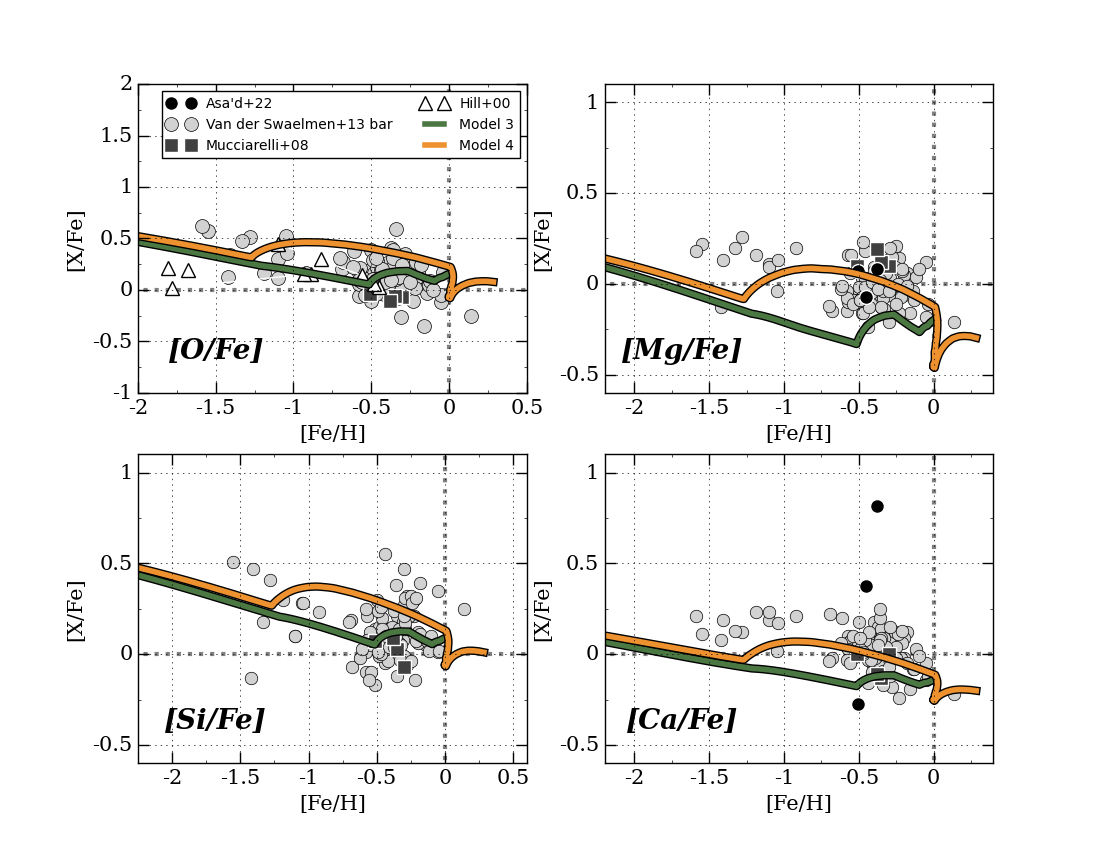}
    \caption{Abundances of four $\alpha$-elements, namely, O, Mg, Si and Ca. The models shown are Model 3 with \citet{HZ09} SFH (green solid line) and Model 4 with \citet{Calura+03} SFH (orange solid line). The data taken from \citet{Asad+22} (black circles), \citet{Mucciarelli+08} (dark grey squares) and \citet{Hill+00} (white triangles) refer to LMC globular clusters whereas those by \citet{vds+13} (light grey circles) regard the LMC bar stars. Since, by means of Table~\ref{tab:constraints} we already excluded Model 1 and Model 2, here we show just the two remaining ones to clarify if we can rule out another one. It can be seen that Model 4 predicts too much Fe abundance, therefore we select Model 3 as our best model.}
    \label{fig:4alfa}
\end{figure*}

Interesting considerations to unravel this ambiguity can come from the analysis of the abundances of the $\alpha$-elements, namely O, Mg, Si and Ca. The nucleosynthesis of these chemical species is known quite well and the stellar yields either have a relatively small uncertainties or, if not, we know how to correct for them. Moreover, the $\alpha$-elements show a very clear behaviour when analysed as a function of [Fe/H]. They generally show a plateau at low [Fe/H] followed by a decrease when [Fe/H] becomes larger. Such a change in slope, that identifies the so called $\alpha$-knee, is important since from it we can infer information regarding the morphological type of the host galaxy and its SFR just based on its position on the [$\alpha$/Fe]-[Fe/H] plane.  In fact, on the basis of the time-delay model (see \citealt{Matteucci21}), the behaviour of the[$\alpha$/Fe]-[Fe/H] relation  and in particular the location of the $\alpha$-knee, depends on the SFH of the specific galaxy taken into exam. The  explanation of that lies in the lifetime of the progenitors of the $\alpha$-elements and of Fe.The $\alpha$-elements are mainly produced by Type II SNe, whereas Type Ia SNe are the most important contributors to Fe abundance. According to stellar physics, Type II SNe  eject their nucleosynthesis products into the ISM on short timescales, on the contrary, Type Ia SNe are characterized by much longer lifetimes. Therefore, at the early stages of galaxy evolution the ISM is polluted mainly by SNeII, which enrich the gas mainly in $\alpha$-elements. They also produce small quantities of Fe, hence [Fe/H] increases slowly. When the first SNe Ia start to explode, injecting large quantities of Fe, the [$\alpha$/Fe] ratio starts decreasing.  If  a galaxy underwent a very mild or gasping star formation, such as irregulars, then the knee is located at very low [Fe/H], whereas in galaxies with more intense star formation such as spirals and spheroids (where the star formation was an intense and short burst) the knee occurs at larger [Fe/H] values.   
The $\alpha$-element behaviour makes it a reliable constraint to chose among different models. 
In Fig.~\ref{fig:4alfa} we show the trend of O, Mg, Si and Ca predicted by Model 3 (Harris \& Zaritsky SFR, green line) and by Model 4 (Calura SFR, yellow line) compared to several sets of data from the literature. In particular we use the LMC globular cluster data from \citet{Hill+00}, \citet{Mucciarelli+08} and \citet{Asad+22}, and the LMC bar stars data by \citet{vds+13}. We can see that the trend is well reproduced by both models and at low [Fe/H] both models are acceptable. Regarding the $\alpha$-knee discussed above, nor the two models neither the data show it in the [Fe/H] range plotted, since it happens earlier given the LMC low level of star formation. The only discriminant factor between Model 3 and Model 4 is the maximum [Fe/H] value reached at the present time. All the four plots show that Model 3 reproduces the data up to the solar [Fe/H] value whereas Model 4 goes further towards a higher Fe content, more similar to the Milky Way one, whereas the data stops earlier. In conclusion, these four plots indicate Model 3, with SFR by \citet{HZ09}, as the best model.
\newline

The last important consideration about Model 3 and Model 4 concerns the way in which the SFRs were derived. \citet{Calura+03} proposed a chemical evolution study of  purely theoretical nature, where the SFR was assumed and compared to the data available at the time (those that we are also showing by \citealt{Hill+00}). The origin of this SFR comes from a previous work by \citet{Bradamante+98} regarding the SFR of dwarf irregulars (and not specifically of the LMC). In that work, they concluded that irregular galaxies should have a SFR characterized by one ore more bursts, for a maximum of 10 bursts, with an efficiency of star formation for each burst from 0.1 Gyr$^{-1}$ to 7 Gyr$^{-1}$. 
The SFR by \citet{HZ09} itself falls within the range proposed by \citet{Bradamante+98}, the only difference is that this one is also supported by observations. As already mentioned in Section~\ref{sec:HZ09}, the authors derived this SFH carrying out a very detailed observational study.  

The last aspect to clarify  about our models is the fraction of systems that can form novae, represented by the factor $\alpha$ in Equation~\ref{eq:nove}. Since our best model is Model 3, the value of the parameter $\alpha$ that can best reproduce the observations is 0.024.

\subsection{\texorpdfstring{$^{26}$}Al and \texorpdfstring{$^{60}$}Fe masses}
\label{sec:al_fe_results}

In this Section we present the results about the masses of $^{26}$Al and $^{60}$Fe in the LMC, as predicted by Model 3.

As mentioned in Section~\ref{sec:yields} we tested two different nova contribution to $^{26}$Al, assuming, at first, no nova production and then testing the yield by \citet{JH07}, whereas for $^{60}$Fe we only have production by massive stars. We report in Table~\ref{tab:alfemass} the results predicted by the four SFHs studied. The first two rows are dedicated to the references where the yields come from, whereas the other four rows list the results about $^{26}$Al and $^{60}$Fe masses. The first column reports the value of the mass of $^{26}$Al obtained excluding the production by nova systems and assuming only a production by massive stars, AGBs and SNIa. The second column contains the mass of $^{26}$Al in the case where also novae are considered among the $^{26}$Al producers. Finally, the third column contains the obtained  mass of $^{60}$Fe. Since for this element there is no contribution coming from novae, there are no multiple cases to discuss.

The values obtained using Model 3 are, as stated since the beginning, predictions for future $\gamma$-observations, thus no comparison with any data can be provided, yet. Some considerations can nevertheless be made, the most important one being the ratio between the mass of the two isotopes. From the observational point of view, we only have data about the Milky Way \citep{Martin+09,Diehl+10,Diehl16}. We know that the instruments detected a much higher flux of $^{26}$Al than that of $^{60}$Fe, and specifically the ratio $^{60}$Fe/$^{26}$Al of the fluxes is around $\sim$15\%.
In \citet{Vasini+22}, where a similar calculation of the masses of $^{26}$Al and $^{60}$Fe was done for the MW, the mass of $^{26}$Al was higher than that of $^{60}$Fe.
 On the contrary, in the case of LMC we obtained the opposite result, predicting an $^{60}$Fe mass of 0.44 M$_{\odot}$, larger than all the  $^{26}$Al values predicted by the four models. 

\begin{table*}
\centering
\begin{tabular}{ccc|c}
    \hline
    & & &\\
    & $^{26}$Al & $^{26}$Al & $^{60}$Fe \\
    & & &\\
    \hline
    massive stars yield & \citet{WW95} & \citet{WW95} & \citet{WW95} \\
    \hline
    nova systems yield & no contribution & \citet{JH07} & no contribution \\
    \hline
    Model 1 & 0.04 M$_{\odot}$ & 0.33 M$_{\odot}$ & 0.13 M$_{\odot}$\\
    \hline
    Model 2 & 0.17 M$_{\odot}$ & 0.57 M$_{\odot}$ & 0.49 M$_{\odot}$\\
    \hline
    Model 3 & 0.20 M$_{\odot}$ & 0.33 M$_{\odot}$ & 0.44 M$_{\odot}$\\
    \hline
    Model 4 & 0.11 M$_{\odot}$ & 0.20 M$_{\odot}$ & 0.29 M$_{\odot}$ \\
    \hline
\end{tabular}
\caption{In the first two rows the references to the yields are listed, whereas from the third to the sixth row we show the results for all our models, with the best one being Model 3. The first column shows the mass of $^{26}$Al produced by massive stars, the second one lists the total mass of $^{26}$Al produced by massive stars together with nova systems and the third column contains the results about $^{60}$Fe.}
\label{tab:alfemass}
\end{table*}

\section{Discussion}
\label{sec:dicussion}
\begin{table*}
\centering
\begin{tabular}{cccc|c}
    \hline
    & & & &\\
    & $^{26}$Al & $^{26}$Al & $^{60}$Fe & $^{60}$Fe/$^{26}$Al \\
    & (massive stars only) & (massive stars + novae) & (massive stars only) & ($^{26}$Al by massive stars \\
    & & & & and novae)\\
    & ph $\cdot$ cm$^{-2}$ s$^{-1}$ & ph $\cdot$ cm $^{-2}$ s$^{-1}$ & ph $\cdot$ cm$^{-2}$ s$^{-1}$ & \\
    & & & &\\
    \hline
    Model 1 & 0.19 $\times 10^{-6}$  & 1.56 $\times 10^{-6}$ & 0.07 $\times 10^{-6}$ & 0.047 \\
    \hline
    Model 2 & 0.8 $\times 10^{-6}$ & 2.7 $\times 10^{-6}$ & 0.28 $\times 10^{-6}$ & 0.10\\
    \hline
    Model 3 & 0.9 $\times 10^{-6}$ & 1.56 $\times 10^{-6}$ & 0.25 $\times 10^{-6}$ & 0.16\\
    \hline
    Model 4 & 0.5 $\times 10^{-6}$ & 0.9 $\times 10^{-6}$ & 0.16 $\times 10^{-6}$ & 0.18\\
    \hline
\end{tabular}
\label{tab:alfe_fluxes}
\caption{Fluxes of $^{26}$Al and $^{60}$Fe for all the models computed using Equation~\ref{eq:flux}. In the first column we present the flux of $^{26}$Al in the case of production only by massive stars, in the second the flux of $^{26}$Al considering also the contribution by novae, in the third we list the flux of $^{60}$Fe. In the fourth we show the flux ratio $^{60}$Fe/$^{26}$Al using the the values from columns 3 and 2.}
\end{table*}
Together with the four models proposed, while carrying out this study we tested some other options. In particular we checked what happens if we assume the same IMF for all the models, namely the \citet{Kroupa1993} one. Adopting this IMF and combining it with the SFRs by \citet{Hasselquist2021}, \citet{HZ09} and \citet{Calura+03} is not totally correct since, as argued in the previous paragraphs, these works assumed specific IMFs, and in order to reproduce those studies we have to make the same assumptions. The results we obtained in these cases did not reproduce the observational constraints of the LMC, thus suggesting that in this galaxy a Salpeter-like IMF should be preferred.
\newline

In Section~\ref{sec:al_fe_results} we presented the masses of these two isotopes predicted only by our best model. We already pointed out that Model 3 predicts more $^{60}$Fe than $^{26}$Al unlike the Milky Way, but more interesting considerations can be made analysing what the other models predicted. In Table~\ref{tab:alfemass} it is noteworthy the fact that both Model 3 and Model 4, namely the two best models, predict more $^{60}$Fe than $^{26}$Al. 
The reason behind this difference of the LMC with respect to the Milky Way lies probably in the IMF.
Both Model 3 and Model 4 adopt a Salpeter IMF whereas Model 1 and Model 2 adopt a Kroupa 93 and a Kroupa 01 IMF respectively. The main producers of both $^{26}$Al and $^{60}$Fe are massive stars, and going to higher and higher masses the yield of $^{60}$Fe is larger than that of $^{26}$Al. The Salpeter IMF predicts a fraction of high mass stars larger than the other two laws, therefore, when adopting a Salpeter IMF we obtain more $^{60}$Fe than $^{26}$Al. 

Regarding the production by nova systems of the $^{26}$Al, we can emphasise that, depending on the considered  model, they contribute to the total fraction in different proportions. For the best model (Model 3) the contribution from novae is $\sim$ 1/3 of the total $^{26}$Al mass. For Model 1, 2 and 4, on the contrary, the fractions are much higher, $\sim$87\%, $\sim$70\% and $\sim$45\% respectively. This percentage can be compared with those obtained in \citet{Vasini+22} for the Milky Way. For the two best models identified in that study, the novae contribute for more than 95\% of the total $^{26}$Al abundance, much more than what they do in the LMC. This is again due to the IMF adopted: since the \citet{Salpeter55} IMF produces more massive stars than the \citet{Kroupa1993} adopted for the Milky Way, the nova systems will contribute less. This explains the discrepancy between the results obtained for the Milky Way and for the LMC.

By adopting the best model (Model 3) we performed two other tests, about different nova yields and electron capture supernovae.
Regarding nova yields, we computed the $^{26}$Al mass making use of one of the most recent set of yields by \citet{Starrfield+2020}. We adopted the three models proposed there, that differ from each other in the chemical composition of the progenitor white dwarf. For each model, they propose yields for different white dwarf masses, that we averaged to obtain one single value for each of the three models. The results show that the contribution by novae with Starrfield yields is much lower, since the yields are lower than those by \citet{JH07}. In the three cases, we obtain 0.205, 0.207 and 0.209 M$_{\odot}$ of $^{26}$Al, which is only slightly more than what we obtain without novae (0.20 M$_{\odot}$).
The second aspect we have investigated is the inclusion of the yields of electron-capture SNe (EC-SNe) by \citet{Wanajo+13} relative to $^{60}$Fe. In particular, we included this class of progenitors assuming three different yields proposed by \citet{Wanajo+13}, 3.61$\times$10$^{-5}$, 7.61$\times$10$^{-5}$ and 13$\times$10$^{-5}$ M$_{\odot}$. In these three cases we obtain 0.45, 0.48 and 0.50 M$_{\odot}$ respectively, whereas without the EC-SN contribution the mass is 0.44 M$_{\odot}$. We can see that EC-SNe increase only slightly the mass of $^{60}$Fe and can be considered negligible.

The last aspect we would like to discuss is related to the uncertainty associated to our results. Concerning the production of $^{26}$Al we can consider the uncertainties in the observed present day nova and star formation rates. By adopting the observed uncertainties listed in Table \ref{tab:constraints} we obtain a minimum mass of $^{26}$Al of $0.20_{-0.07}^{+0.09}$ M$_{\odot}$ and a maximum mass of $0.33_{-0.04}^{+0.27}$ M$_{\odot}$. Given these values we predict a mass of $^{26}$Al in the range 0.13 -- 0.60 M$_{\odot}$.
By repeating the same procedure for $^{60}$Fe, we predict its mass to be $0.44_{-0.18}^{+0.76}$ M$_{\odot}$.

In the light of the upcoming COSI mission in 2027, these predicted radioactive masses can be converted to quasi-steady-state $\gamma$-ray line fluxes.
Assuming a point-like distribution at a distance $d_{\rm LMC} = 49.6$\,kpc to the LMC as a first order estimate, the $\gamma$-ray line flux $F_i$ from a radioactive element $i$ can be calculated as:
\begin{equation}
	F_i = \frac{M_i p_i}{m_i \tau_i} \frac{1}{4 \pi d_{\rm LMC}^2}\mathrm{,}
\label{eq:flux}
\end{equation}
where $M_i$ is the mass of element $i$ that we predicted, $m_i$ its isotopic mass, $\tau_i$ its lifetime, and $p_i$ the probability to emit a $\gamma$-ray photon after the decay.
For $p \approx 1$ for both isotopes, the flux at 1.809\,MeV from the decay of $^{26}$Al is $1.6 \times 10^{-6}\,\mathrm{ph\,cm^{-2}\,s^{-1}}$, and at 1.173 and 1.332\,MeV from the decay of $^{60}$Fe is $2.5 \times 10^{-7}\,\mathrm{ph\,cm^{-2}\,s^{-1}}$, respectively, for Model 3.
We summarise the flux estimated from our models in Table~\ref{tab:alfe_fluxes}. The first column contains the predicted flux of $^{26}$Al in case of production only by massive stars, the second one refers to the case of $^{26}$Al produced by massive stars together with novae, the third shows the $^{60}$Fe flux and the fourth lists the flux ratio between column 3 and column 2.
If the fluxes are distributed across the actual size of the LMC on the sky of about $4^{\circ} \times 5^{\circ}$, the extended emission lowers the expected total fluxes by less than 5\,\%, but the flux per resolution element \citep{Tomsick+19} is then about 4--6 times smaller, requiring a somewhat longer integration time than the nominal two-year mission.
The exact exposure time required to detect the 1.809\,MeV signal from the LMC then depends on the actual morphology of the galaxy, that is, where the emission could be found.
Given the HI measurements of \citep{Dawson+13} of the LMC, a superbubble structure is evident which may concentrate more $^{26}$Al in fewer cavities than being spread out across the entire galaxy.
In the Milky Way, $^{26}$Al is also found inside superbubbles (e.g. \citealt{Diehl+10,SD17,Krause+15}), so that the same idea for the LMC is not unreasonable.
For our best model, then, the LMC could be detectable in $^{26}$Al $\gamma$-rays within 2–4 years with COSI.

In the case of $^{60}$Fe, albeit having a larger persistent mass, the lifetime is longer and the nuclei masses larger, so that the fluxes for each of the two decay lines at 1.173 and 1.332\,MeV, respectively, are about a factor of 8 smaller than the $^{26}$Al line.
This means that even with COSI, $^{60}$Fe $\gamma$-rays will not be detected from the LMC unless additional contributions from, as of yet unknown $^{60}$Fe producers, would be enhanced in this galaxy.

\section{Conclusions}
\label{sec:conclusions}

In this work we developed a model to study the chemical evolution of two radioactive isotopes, $^{26}$Al and $^{60}$Fe, in the Large Magellanic Cloud. We tested four different cases with different star formation histories taken from the literature and we selected as best model, the one that could reproduce at best the present time LMC observations and $\alpha$-elements abundance patterns. Our results can be summarized as follows:

\begin{itemize}

    \item according to the tests we performed, the model that can best reproduce the present time features of the LMC is Model 3, that adopts a star formation history by \citet{HZ09} and an initial mass function by \citet{Salpeter55}.  These assumptions allow us to reproduce the present time SFR, SNe rate, nova rate, gas mass and metallicity. Moreover, they also fit the abundances of the most important $\alpha$-elements such as O, Mg, Si, and Ca. In general, this indicates that the LMC underwent a very low star formation at the beginning of its history, followed by a recent burst that is still active. This behaviour agrees with the characteristics of an irregular galaxy, as the LMC is;
    
    \item our best model, Model 3, predicts $\sim$0.44 M$_{\odot}$ of $^{60}$Fe and a $^{26}$Al mass in the range 0.20 -- 0.33 M$_{\odot}$, depending on the presence or absence of the nova contribution. However, considering the uncertainties observed in nova and in star formation rates, we predict a $^{26}$Al mass in the range 0.13 -- 0.60 M$_{\odot}$ and a $^{60}$Fe mass of  $0.44_{-0.18}^{+0.76}$ M$_{\odot}$. Even though the intervals partially overlap, the models always predict more $^{60}$Fe than $^{26}$Al. This is in contrast with what we found with a similar modelling approach for the Milky Way, where the mass of $^{60}$Fe is smaller than that of $^{26}$Al. This result is reinforced by the fact that also Model 4, that is the second most reliable model from this study, predicts more $^{60}$Fe than $^{26}$Al. The reason for this lies in the Salpeter IMF adopted by these two models: in fact, this IMF predicts more massive stars than the IMF suitable for the MW disk. Since these stars are the main producers of $^{60}$Fe, then $^{60}$Fe abundance is larger than $^{26}$Al abundance. It is important to highlight that the uncertainties in the model and in the input parameters, such as the yields, are not the main cause of the discrepancy between LMC and MW, simply because we adopted the same chemical evolution model with the same yields for the two galaxies. Therefore, this can be interpreted as an evidence to prefer a Salpeter-like IMF when describing the LMC;

    \item according to our best model, the contribution to $^{26}$Al coming from nova systems should not exceed 1/3 of the total abundance. This percentage is much lower than what obtained in \citet{Vasini+22} for the Milky Way, where it was $\sim$95\% of the total $^{26}$Al abundance. Again, this is mainly due to the larger fraction of massive stars in the \citet{Salpeter55} IMF relative to the \citet{Kroupa1993} one;

    \item given the estimated sensitivity of the upcoming COSI-SMEX mission in 2027, 1.809\,MeV $\gamma$-rays from the decay of $^{26}$Al at a flux level of $(1$--$2) \times 10^{-6}\,\mathrm{ph\,cm^{-2}\,s^{-1}}$ is within reach. The $\gamma$-ray lines from the decay of $^{60}$Fe at 1.173 and 1.332\,MeV are probably too faint to be detectable from the LMC, even with improved instrumentation.

\end{itemize}

\section*{Acknowledgements}

The authors kindly thank the referee F.-K. Thielemann for the interesting comments and suggestions that improved the paper.
A. Vasini and F. Matteucci thank I.N.A.F. for the 1.05.12.06.05 Theory Grant - Galactic archaeology with radioactive and stable nuclei.
E. Spitoni  received funding from the European Union’s Horizon 2020 research and innovation program under SPACE-H2020 grant agreement number 101004214 (EXPLORE project). 
\section*{Data Availability}
The data showed in this article will be shared on reasonable request to the corresponding author.


\bibliographystyle{mnras}
\bibliography{example} 

\begin{thebibliography}{}
\makeatletter
\relax
\def\mn@urlcharsother{\let\do\@makeother \do\$\do\&\do\#\do\^\do\_\do\%\do\~}
\def\mn@doi{\begingroup\mn@urlcharsother \@ifnextchar [ {\mn@doi@}
  {\mn@doi@[]}}
\def\mn@doi@[#1]#2{\def\@tempa{#1}\ifx\@tempa\@empty \href
  {http://dx.doi.org/#2} {doi:#2}\else \href {http://dx.doi.org/#2} {#1}\fi
  \endgroup}
\def\mn@eprint#1#2{\mn@eprint@#1:#2::\@nil}
\def\mn@eprint@arXiv#1{\href {http://arxiv.org/abs/#1} {{\tt arXiv:#1}}}
\def\mn@eprint@dblp#1{\href {http://dblp.uni-trier.de/rec/bibtex/#1.xml}
  {dblp:#1}}
\def\mn@eprint@#1:#2:#3:#4\@nil{\def\@tempa {#1}\def\@tempb {#2}\def\@tempc
  {#3}\ifx \@tempc \@empty \let \@tempc \@tempb \let \@tempb \@tempa \fi \ifx
  \@tempb \@empty \def\@tempb {arXiv}\fi \@ifundefined
  {mn@eprint@\@tempb}{\@tempb:\@tempc}{\expandafter \expandafter \csname
  mn@eprint@\@tempb\endcsname \expandafter{\@tempc}}}

\bibitem[\protect\citeauthoryear{{Asa'd}, {Hernandez}, {As'ad}, {Molero},
  {Matteucci}, {Larsen}  \& {Chilingarian}}{{Asa'd} et~al.}{2022}]{Asad+22}
{Asa'd} R.,  {Hernandez} S.,  {As'ad} A.,  {Molero} M.,  {Matteucci} F.,
  {Larsen} S.,   {Chilingarian} I.~V.,  2022, \mn@doi [\apj]
  {10.3847/1538-4357/ac5f3e}, \href
  {https://ui.adsabs.harvard.edu/abs/2022ApJ...929..174A} {929, 174}

\bibitem[\protect\citeauthoryear{{Bradamante}, {Matteucci}  \&
  {D'Ercole}}{{Bradamante} et~al.}{1998}]{Bradamante+98}
{Bradamante} F.,  {Matteucci} F.,   {D'Ercole} A.,  1998, \mn@doi [\aap]
  {10.48550/arXiv.astro-ph/9801131}, \href
  {https://ui.adsabs.harvard.edu/abs/1998A&A...337..338B} {337, 338}

\bibitem[\protect\citeauthoryear{{Calura}, {Matteucci}  \& {Vladilo}}{{Calura}
  et~al.}{2003}]{Calura+03}
{Calura} F.,  {Matteucci} F.,   {Vladilo} G.,  2003, \mn@doi [\mnras]
  {10.1046/j.1365-8711.2003.06197.x}, \href
  {https://ui.adsabs.harvard.edu/abs/2003MNRAS.340...59C} {340, 59}

\bibitem[\protect\citeauthoryear{{Clayton}}{{Clayton}}{1984}]{Clayton84}
{Clayton} D.~D.,  1984, \mn@doi [\apj] {10.1086/162518}, \href
  {https://ui.adsabs.harvard.edu/abs/1984ApJ...285..411C} {285, 411}

\bibitem[\protect\citeauthoryear{{Clayton}}{{Clayton}}{1988}]{Clayton88}
{Clayton} D.~D.,  1988, \mn@doi [\mnras] {10.1093/mnras/234.1.1}, \href
  {https://ui.adsabs.harvard.edu/abs/1988MNRAS.234....1C} {234, 1}

\bibitem[\protect\citeauthoryear{{D'Antona} \& {Matteucci}}{{D'Antona} \&
  {Matteucci}}{1991}]{D'antona&Matteucci91}
{D'Antona} F.,  {Matteucci} F.,  1991, \aap, \href
  {https://ui.adsabs.harvard.edu/abs/1991A&A...248...62D} {248, 62}

\bibitem[\protect\citeauthoryear{{Dawson}, {McClure-Griffiths}, {Wong},
  {Dickey}, {Hughes}, {Fukui}  \& {Kawamura}}{{Dawson}
  et~al.}{2013}]{Dawson+13}
{Dawson} J.~R.,  {McClure-Griffiths} N.~M.,  {Wong} T.,  {Dickey} J.~M.,
  {Hughes} A.,  {Fukui} Y.,   {Kawamura} A.,  2013, \mn@doi [\apj]
  {10.1088/0004-637X/763/1/56}, \href
  {https://ui.adsabs.harvard.edu/abs/2013ApJ...763...56D} {763, 56}

\bibitem[\protect\citeauthoryear{{Diehl}}{{Diehl}}{2013}]{Diehl13}
{Diehl} R.,  2013, \mn@doi [Reports on Progress in Physics]
  {10.1088/0034-4885/76/2/026301}, \href
  {https://ui.adsabs.harvard.edu/abs/2013RPPh...76b6301D} {76, 026301}

\bibitem[\protect\citeauthoryear{{Diehl}}{{Diehl}}{2016}]{Diehl16}
{Diehl} R.,  2016, in Journal of Physics Conference Series. p. 012011,
  \mn@doi{10.1088/1742-6596/665/1/012011}

\bibitem[\protect\citeauthoryear{{Diehl} et~al.,}{{Diehl}
  et~al.}{1995}]{Diehl+95}
{Diehl} R.,  et~al., 1995, \aap, \href
  {https://ui.adsabs.harvard.edu/abs/1995A&A...298..445D} {298, 445}

\bibitem[\protect\citeauthoryear{{Diehl} et~al.,}{{Diehl}
  et~al.}{2006}]{Diehl+06}
{Diehl} R.,  et~al., 2006, \mn@doi [\nat] {10.1038/nature04364}, \href
  {https://ui.adsabs.harvard.edu/abs/2006Natur.439...45D} {439, 45}

\bibitem[\protect\citeauthoryear{{Diehl} et~al.,}{{Diehl}
  et~al.}{2010}]{Diehl+10}
{Diehl} R.,  et~al., 2010, \mn@doi [\aap] {10.1051/0004-6361/201014302}, \href
  {https://ui.adsabs.harvard.edu/abs/2010A&A...522A..51D} {522, A51}

\bibitem[\protect\citeauthoryear{{Ford}}{{Ford}}{1978}]{Ford+78}
{Ford} H.~C.,  1978, \mn@doi [\apj] {10.1086/155819}, \href
  {https://ui.adsabs.harvard.edu/abs/1978ApJ...219..595F} {219, 595}

\bibitem[\protect\citeauthoryear{{Harris} \& {Zaritsky}}{{Harris} \&
  {Zaritsky}}{2009}]{HZ09}
{Harris} J.,  {Zaritsky} D.,  2009, \mn@doi [\aj]
  {10.1088/0004-6256/138/5/1243}, \href
  {https://ui.adsabs.harvard.edu/abs/2009AJ....138.1243H} {138, 1243}

\bibitem[\protect\citeauthoryear{{Hasselquist} et~al.,}{{Hasselquist}
  et~al.}{2021}]{Hasselquist2021}
{Hasselquist} S.,  et~al., 2021, \mn@doi [\apj] {10.3847/1538-4357/ac25f9},
  \href {https://ui.adsabs.harvard.edu/abs/2021ApJ...923..172H} {923, 172}

\bibitem[\protect\citeauthoryear{{Hill}, {Fran{\c{c}}ois}, {Spite}, {Primas}
  \& {Spite}}{{Hill} et~al.}{2000}]{Hill+00}
{Hill} V.,  {Fran{\c{c}}ois} P.,  {Spite} M.,  {Primas} F.,   {Spite} F.,
  2000, \mn@doi [\aap] {10.48550/arXiv.astro-ph/0009273}, \href
  {https://ui.adsabs.harvard.edu/abs/2000A&A...364L..19H} {364, L19}

\bibitem[\protect\citeauthoryear{{Iwamoto}, {Brachwitz}, {Nomoto}, {Kishimoto},
  {Umeda}, {Hix}  \& {Thielemann}}{{Iwamoto} et~al.}{1999}]{Iwamoto+99}
{Iwamoto} K.,  {Brachwitz} F.,  {Nomoto} K.,  {Kishimoto} N.,  {Umeda} H.,
  {Hix} W.~R.,   {Thielemann} F.-K.,  1999, \mn@doi [\apjs] {10.1086/313278},
  \href {https://ui.adsabs.harvard.edu/abs/1999ApJS..125..439I} {125, 439}

\bibitem[\protect\citeauthoryear{{Jos{\'e}} \& {Hernanz}}{{Jos{\'e}} \&
  {Hernanz}}{1998}]{JH98}
{Jos{\'e}} J.,  {Hernanz} M.,  1998, \mn@doi [\apj] {10.1086/305244}, \href
  {https://ui.adsabs.harvard.edu/abs/1998ApJ...494..680J} {494, 680}

\bibitem[\protect\citeauthoryear{{Jos{\'e}} \& {Hernanz}}{{Jos{\'e}} \&
  {Hernanz}}{2007}]{JH07}
{Jos{\'e}} J.,  {Hernanz} M.,  2007, \mn@doi [Journal of Physics G Nuclear
  Physics] {10.1088/0954-3899/34/12/R01}, \href
  {https://ui.adsabs.harvard.edu/abs/2007JPhG...34..431J} {34, R431}

\bibitem[\protect\citeauthoryear{{Karakas}}{{Karakas}}{2010}]{Karakas10}
{Karakas} A.~I.,  2010, \mn@doi [\mnras] {10.1111/j.1365-2966.2009.16198.x},
  \href {https://ui.adsabs.harvard.edu/abs/2010MNRAS.403.1413K} {403, 1413}

\bibitem[\protect\citeauthoryear{{Kennicutt}}{{Kennicutt}}{1998}]{Kennicutt1998}
{Kennicutt} Robert~C. J.,  1998, \mn@doi [\apj] {10.1086/305588}, \href
  {https://ui.adsabs.harvard.edu/abs/1998ApJ...498..541K} {498, 541}

\bibitem[\protect\citeauthoryear{{Kobayashi}, {Umeda}, {Nomoto}, {Tominaga}  \&
  {Ohkubo}}{{Kobayashi} et~al.}{2006}]{Kobayashi+06}
{Kobayashi} C.,  {Umeda} H.,  {Nomoto} K.,  {Tominaga} N.,   {Ohkubo} T.,
  2006, \mn@doi [\apj] {10.1086/508914}, \href
  {https://ui.adsabs.harvard.edu/abs/2006ApJ...653.1145K} {653, 1145}

\bibitem[\protect\citeauthoryear{{Kobayashi}, {Karakas}  \&
  {Umeda}}{{Kobayashi} et~al.}{2011}]{Kobayashi+11}
{Kobayashi} C.,  {Karakas} A.~I.,   {Umeda} H.,  2011, \mn@doi [\mnras]
  {10.1111/j.1365-2966.2011.18621.x}, \href
  {https://ui.adsabs.harvard.edu/abs/2011MNRAS.414.3231K} {414, 3231}

\bibitem[\protect\citeauthoryear{{Krause} et~al.,}{{Krause}
  et~al.}{2015}]{Krause+15}
{Krause} M. G.~H.,  et~al., 2015, \mn@doi [\aap] {10.1051/0004-6361/201525847},
  \href {https://ui.adsabs.harvard.edu/abs/2015A&A...578A.113K} {578, A113}

\bibitem[\protect\citeauthoryear{{Kretschmer}, {Diehl}, {Krause}, {Burkert},
  {Fierlinger}, {Gerhard}, {Greiner}  \& {Wang}}{{Kretschmer}
  et~al.}{2013}]{Kretschmer+13}
{Kretschmer} K.,  {Diehl} R.,  {Krause} M.,  {Burkert} A.,  {Fierlinger} K.,
  {Gerhard} O.,  {Greiner} J.,   {Wang} W.,  2013, \mn@doi [\aap]
  {10.1051/0004-6361/201322563}, \href
  {https://ui.adsabs.harvard.edu/abs/2013A&A...559A..99K} {559, A99}

\bibitem[\protect\citeauthoryear{{Kroupa}}{{Kroupa}}{2001}]{Kroupa01}
{Kroupa} P.,  2001, \mn@doi [\mnras] {10.1046/j.1365-8711.2001.04022.x}, \href
  {https://ui.adsabs.harvard.edu/abs/2001MNRAS.322..231K} {322, 231}

\bibitem[\protect\citeauthoryear{{Kroupa}, {Tout}  \& {Gilmore}}{{Kroupa}
  et~al.}{1993}]{Kroupa1993}
{Kroupa} P.,  {Tout} C.~A.,   {Gilmore} G.,  1993, \mn@doi [\mnras]
  {10.1093/mnras/262.3.545}, \href
  {https://ui.adsabs.harvard.edu/abs/1993MNRAS.262..545K} {262, 545}

\bibitem[\protect\citeauthoryear{{Limongi} \& {Chieffi}}{{Limongi} \&
  {Chieffi}}{2006}]{LC06}
{Limongi} M.,  {Chieffi} A.,  2006, \mn@doi [\apj] {10.1086/505164}, \href
  {https://ui.adsabs.harvard.edu/abs/2006ApJ...647..483L} {647, 483}

\bibitem[\protect\citeauthoryear{{Limongi} \& {Chieffi}}{{Limongi} \&
  {Chieffi}}{2018}]{LC18}
{Limongi} M.,  {Chieffi} A.,  2018, \mn@doi [\apjs] {10.3847/1538-4365/aacb24},
  \href {https://ui.adsabs.harvard.edu/abs/2018ApJS..237...13L} {237, 13}

\bibitem[\protect\citeauthoryear{{Mannucci}, {Della Valle}, {Panagia},
  {Cappellaro}, {Cresci}, {Maiolino}, {Petrosian}  \& {Turatto}}{{Mannucci}
  et~al.}{2005}]{Mannucci+05}
{Mannucci} F.,  {Della Valle} M.,  {Panagia} N.,  {Cappellaro} E.,  {Cresci}
  G.,  {Maiolino} R.,  {Petrosian} A.,   {Turatto} M.,  2005, \mn@doi [\aap]
  {10.1051/0004-6361:20041411}, \href
  {https://ui.adsabs.harvard.edu/abs/2005A&A...433..807M} {433, 807}

\bibitem[\protect\citeauthoryear{{Maoz} \& {Badenes}}{{Maoz} \&
  {Badenes}}{2010}]{Maoz+10}
{Maoz} D.,  {Badenes} C.,  2010, \mn@doi [\mnras]
  {10.1111/j.1365-2966.2010.16988.x}, \href
  {https://ui.adsabs.harvard.edu/abs/2010MNRAS.407.1314M} {407, 1314}

\bibitem[\protect\citeauthoryear{{Martin}, {Kn{\"o}dlseder}, {Vink},
  {Decourchelle}  \& {Renaud}}{{Martin} et~al.}{2009}]{Martin+09}
{Martin} P.,  {Kn{\"o}dlseder} J.,  {Vink} J.,  {Decourchelle} A.,   {Renaud}
  M.,  2009, \mn@doi [\aap] {10.1051/0004-6361/200809735}, \href
  {https://ui.adsabs.harvard.edu/abs/2009A&A...502..131M} {502, 131}

\bibitem[\protect\citeauthoryear{{Matteucci}}{{Matteucci}}{2021}]{Matteucci21}
{Matteucci} F.,  2021, \mn@doi [\aapr] {10.1007/s00159-021-00133-8}, \href
  {https://ui.adsabs.harvard.edu/abs/2021A&ARv..29....5M} {29, 5}

\bibitem[\protect\citeauthoryear{{Matteucci} \& {Chiosi}}{{Matteucci} \&
  {Chiosi}}{1983}]{MC83}
{Matteucci} F.,  {Chiosi} C.,  1983, \aap, \href
  {https://ui.adsabs.harvard.edu/abs/1983A&A...123..121M} {123, 121}

\bibitem[\protect\citeauthoryear{{Matteucci} \& {Greggio}}{{Matteucci} \&
  {Greggio}}{1986}]{Matteucci&Greggio86}
{Matteucci} F.,  {Greggio} L.,  1986, \aap, \href
  {https://ui.adsabs.harvard.edu/abs/1986A&A...154..279M} {154, 279}

\bibitem[\protect\citeauthoryear{{Matteucci} \& {Recchi}}{{Matteucci} \&
  {Recchi}}{2001}]{Matteucci&Recchi01}
{Matteucci} F.,  {Recchi} S.,  2001, \mn@doi [\apj] {10.1086/322472}, \href
  {https://ui.adsabs.harvard.edu/abs/2001ApJ...558..351M} {558, 351}

\bibitem[\protect\citeauthoryear{{Mr{\'o}z} et~al.,}{{Mr{\'o}z}
  et~al.}{2016}]{Mroz+16}
{Mr{\'o}z} P.,  et~al., 2016, \mn@doi [\apjs] {10.3847/0067-0049/222/1/9},
  \href {https://ui.adsabs.harvard.edu/abs/2016ApJS..222....9M} {222, 9}

\bibitem[\protect\citeauthoryear{{Mucciarelli}, {Carretta}, {Origlia}  \&
  {Ferraro}}{{Mucciarelli} et~al.}{2008}]{Mucciarelli+08}
{Mucciarelli} A.,  {Carretta} E.,  {Origlia} L.,   {Ferraro} F.~R.,  2008,
  \mn@doi [\aj] {10.1088/0004-6256/136/1/375}, \href
  {https://ui.adsabs.harvard.edu/abs/2008AJ....136..375M} {136, 375}

\bibitem[\protect\citeauthoryear{{Nidever} et~al.,}{{Nidever}
  et~al.}{2020}]{Nidever+20}
{Nidever} D.~L.,  et~al., 2020, \mn@doi [\apj] {10.3847/1538-4357/ab7305},
  \href {https://ui.adsabs.harvard.edu/abs/2020ApJ...895...88N} {895, 88}

\bibitem[\protect\citeauthoryear{{Nosowitz} et~al.,}{{Nosowitz}
  et~al.}{2022}]{Nosowitz+22}
{Nosowitz} J.,  et~al., 2022, in American Astronomical Society Meeting
  Abstracts. p. 103.02

\bibitem[\protect\citeauthoryear{{Pleintinger}, {Diehl}, {Siegert}, {Greiner}
  \& {Krause}}{{Pleintinger} et~al.}{2022}]{Pleintinger+20}
{Pleintinger} M. M.~M.,  {Diehl} R.,  {Siegert} T.,  {Greiner} J.,   {Krause}
  M. G.~H.,  2022, \mn@doi [arXiv e-prints] {10.48550/arXiv.2212.11228}, \href
  {https://ui.adsabs.harvard.edu/abs/2022arXiv221211228P} {p. arXiv:2212.11228}

\bibitem[\protect\citeauthoryear{{Romano} \& {Matteucci}}{{Romano} \&
  {Matteucci}}{2003}]{Romano&Matteucci03}
{Romano} D.,  {Matteucci} F.,  2003, \mn@doi [\mnras]
  {10.1046/j.1365-8711.2003.06526.x}, \href
  {https://ui.adsabs.harvard.edu/abs/2003MNRAS.342..185R} {342, 185}

\bibitem[\protect\citeauthoryear{{Salpeter}}{{Salpeter}}{1955}]{Salpeter55}
{Salpeter} E.~E.,  1955, \mn@doi [\apj] {10.1086/145971}, \href
  {https://ui.adsabs.harvard.edu/abs/1955ApJ...121..161S} {121, 161}

\bibitem[\protect\citeauthoryear{{Siegert} \& {Diehl}}{{Siegert} \&
  {Diehl}}{2017}]{SD17}
{Siegert} T.,  {Diehl} R.,  2017, in {Kubono} S.,  {Kajino} T.,  {Nishimura}
  S.,  {Isobe} T.,  {Nagataki} S.,  {Shima} T.,   {Takeda} Y.,  eds, 14th
  International Symposium on Nuclei in the Cosmos (NIC2016). p. 020305
  (\mn@eprint {arXiv} {1609.08817}), \mn@doi{10.7566/JPSCP.14.020305}

\bibitem[\protect\citeauthoryear{{Starrfield}, {Bose}, {Iliadis}, {Hix},
  {Woodward}  \& {Wagner}}{{Starrfield} et~al.}{2020}]{Starrfield+2020}
{Starrfield} S.,  {Bose} M.,  {Iliadis} C.,  {Hix} W.~R.,  {Woodward} C.~E.,
  {Wagner} R.~M.,  2020, \mn@doi [\apj] {10.3847/1538-4357/ab8d23}, \href
  {https://ui.adsabs.harvard.edu/abs/2020ApJ...895...70S} {895, 70}

\bibitem[\protect\citeauthoryear{{Tchernyshyov}, {Meixner}, {Seale}, {Fox},
  {Friedman}, {Dwek}  \& {Galliano}}{{Tchernyshyov}
  et~al.}{2015}]{Tchernyshyov+15}
{Tchernyshyov} K.,  {Meixner} M.,  {Seale} J.,  {Fox} A.,  {Friedman} S.~D.,
  {Dwek} E.,   {Galliano} F.,  2015, \mn@doi [\apj]
  {10.1088/0004-637X/811/2/78}, \href
  {https://ui.adsabs.harvard.edu/abs/2015ApJ...811...78T} {811, 78}

\bibitem[\protect\citeauthoryear{{Timmes}, {Woosley}, {Hartmann}, {Hoffman},
  {Weaver}  \& {Matteucci}}{{Timmes} et~al.}{1995}]{Timmes+95}
{Timmes} F.~X.,  {Woosley} S.~E.,  {Hartmann} D.~H.,  {Hoffman} R.~D.,
  {Weaver} T.~A.,   {Matteucci} F.,  1995, \mn@doi [\apj] {10.1086/176046},
  \href {https://ui.adsabs.harvard.edu/abs/1995ApJ...449..204T} {449, 204}

\bibitem[\protect\citeauthoryear{{Tomsick} et~al.,}{{Tomsick}
  et~al.}{2019}]{Tomsick+19}
{Tomsick} J.,  et~al., 2019, in Bulletin of the American Astronomical Society.
  p.~98 (\mn@eprint {arXiv} {1908.04334}), \mn@doi{10.48550/arXiv.1908.04334}

\bibitem[\protect\citeauthoryear{{Van der Swaelmen}, {Hill}, {Primas}  \&
  {Cole}}{{Van der Swaelmen} et~al.}{2013}]{vds+13}
{Van der Swaelmen} M.,  {Hill} V.,  {Primas} F.,   {Cole} A.~A.,  2013, \mn@doi
  [\aap] {10.1051/0004-6361/201321109}, \href
  {https://ui.adsabs.harvard.edu/abs/2013A&A...560A..44V} {560, A44}

\bibitem[\protect\citeauthoryear{{Vasini}, {Matteucci}  \& {Spitoni}}{{Vasini}
  et~al.}{2022}]{Vasini+22}
{Vasini} A.,  {Matteucci} F.,   {Spitoni} E.,  2022, \mn@doi [\mnras]
  {10.1093/mnras/stac2981}, \href
  {https://ui.adsabs.harvard.edu/abs/2022MNRAS.517.4256V} {517, 4256}

\bibitem[\protect\citeauthoryear{{Vedrenne} et~al.,}{{Vedrenne}
  et~al.}{2003}]{Vedrenne+03}
{Vedrenne} G.,  et~al., 2003, \mn@doi [\aap] {10.1051/0004-6361:20031482},
  \href {https://ui.adsabs.harvard.edu/abs/2003A&A...411L..63V} {411, L63}

\bibitem[\protect\citeauthoryear{{Wanajo}, {Janka}  \& {M{\"u}ller}}{{Wanajo}
  et~al.}{2013}]{Wanajo+13}
{Wanajo} S.,  {Janka} H.-T.,   {M{\"u}ller} B.,  2013, \mn@doi [\apjl]
  {10.1088/2041-8205/774/1/L6}, \href
  {https://ui.adsabs.harvard.edu/abs/2013ApJ...774L...6W} {774, L6}

\bibitem[\protect\citeauthoryear{{Winkler} et~al.,}{{Winkler}
  et~al.}{2003}]{Winkler+03}
{Winkler} C.,  et~al., 2003, \mn@doi [\aap] {10.1051/0004-6361:20031288}, \href
  {https://ui.adsabs.harvard.edu/abs/2003A&A...411L...1W} {411, L1}

\bibitem[\protect\citeauthoryear{{Woosley} \& {Weaver}}{{Woosley} \&
  {Weaver}}{1995}]{WW95}
{Woosley} S.~E.,  {Weaver} T.~A.,  1995, \mn@doi [\apjs] {10.1086/192237},
  \href {https://ui.adsabs.harvard.edu/abs/1995ApJS..101..181W} {101, 181}

\makeatother
\end{thebibliography}




\bsp	
\label{lastpage}
\end{document}